\documentstyle[12pt,epsfig]{article}
\begin{document}
\input psfig

\title{\bf Breakup probabilities and the optical potential in elastic
 deuteron scattering \vspace{20mm}}
\author{A. Ingemarsson$^{1}$ and R. Shyam$^{2}$ \\[1ex]
$^{1}$ {The Svedberg Laboratory and Department of Radiation Sciences,}\\ 
 {Box 533, S-75121 Uppsala, Sweden} \\
 {e-mail Anders.Ingemarsson@tsl.uu.se} \\[1ex]
$^{2}$	{Saha Institute of Nuclear Physics,}\\
{ 1/AF Bidhan Nagar, Calcutta - 700064, India}\\
  {e-mail Shyam@tnp.saha.ernet.in}} 

\maketitle
\vspace{5mm}
\begin{abstract}
\baselineskip  
 24pt
We investigate the effect of the deuteron breakup on the optical potential
for the elastic scattering of 56 MeV deuterons from $^{51}$V.
The breakup probabilities 
calculated within the post-form distorted wave Born-approximation theory of
breakup reactions are fitted to derive the contribution
to the optical potential from the breakup channels. The breakup potentials 
are found to have large real as well as imaginary parts. Thus the dynamical
polarization potential due to the breakup process is
expected to modify strongly the real part of the optical potential
calculated by the double folding model. 
\end{abstract}
\newpage

\section{Introduction}
\baselineskip 
 24pt
The folding model with realistic effective nucleon-nucleon interactions
has given a good insight into the nucleus-nucleus interaction (see e.g. a 
recent review by Brandan and Satchler~\cite{Bra96}). However,  
when a light ion interacts with other ions there is also a
possibility of the breakup of the projectile into two or more
fragments. If the breakup channel is strong, 
it will affect not only the imaginary potential but also the
real one. This leads to a dynamical polarisation potential
(DPP) which has to be added to the real potential obtained by the folding
model. The DPP due to the breakup process can,  
for example, be estimated from the adiabatic model proposed 
by Johnson and Soper~\cite{John70}, or by more sophisticated
continuum discretised coupled channels techniques~\cite{Sak86}. The DPP 
is required to describe the elastic scattering of
the weakly bound projectiles like $^{6,7}$Li and $^{9}$Be in the folding
model with reasonable renormalisation factors ~\cite{Bra96}.

Experimental studies have shown that the breakup probabilities
increase drastically with energy even for tightly bound projectiles 
~\cite{Wu78, Bud78}. For example, the cross section for the  
breakup of the $\alpha$ particle increases by, at least, an order of
magnitude as the beam energy is varied from 65 MeV to
140 MeV~\cite{Mei83,Koo79}. At higher beam energies ($\geq$ 140 MeV)
the breakup cross section can be as large as 25$\%$
of the total reaction cross section. Thus the DPP could be required
also for tightly bound projectiles for beam energies above 30 MeV/A.

In a recent report~\cite{Ing96} one of us investigated the contributions
to the $\alpha$-particle optical potential from the ($\alpha,^{3}$He) 
breakup reaction on $^{62}$Ni target at the incident energy of 172.5 MeV.
The calculations proceed in two steps. First 
the breakup probabilities are calculated within a theory
which is formulated in the frame-work of the post form distorted-wave
Born-approximation (PFDWBA).
This theory has been found to reproduce the experimental data on the
breakup of light projectiles extremely well~\cite{Shyam84}.  
In the second step, these probabilities are fitted to generate the breakup
part of the optical potential which is assumed to consist of two parts; one
which is entirely due to the breakup of the projectile and the another
independent of it. It was found in ~\cite{Ing96} that breakup   
contributes substantially  to the optical potential.
However, in ~\cite{Ing96}, the DPP for the alpha elastic scattering was 
obtained by fitting only to the $(\alpha,^3$He) breakup channel.
For a complete determination of the DPP due to the breakup process 
the breakup probabilities for all the $\alpha$-breakup channels, namely, 
($\alpha$,p), ($\alpha$,n), ($\alpha$,d), 
($\alpha,^{3}$He) and ($\alpha$,t) should be calculated and fitted.

Before taking up this rather ambitious task, which nobody has investigated
so far, we considered it worthwhile to perform calculations for the
simple system of the deuteron to test the method in detail. In this
case there are only two breakup channels, $(d,p)$ and $(d,n)$ and
the experimental data on the deuteron breakup (see e.g.~\cite{Mat80})
is quite comprehensive.  In this paper, we investigate the scattering
of 56 MeV deuterons from $^{51}$V. The breakup probabilities for the
$(d,p)$ and $(d,n)$ channels are calculated for this particular case.
These are fitted to determine the DPP due to the breakup process. 

In sections 2 we describe the calculation of the breakup  
probabilities in the optical model. The calculation of the same within the
PFDWBA theory of breakup is presented in section 3.
The results for the DPP obtained by fitting the total and elastic breakup
probabilities are discussed in sections 4.  Our conclusions are
presented in section 5.
 
\section{Calculations of breakup probabilities from the optical model.}

The calculations are based on the same principles as in Ref.~\cite{Ing96}.
We assume that the optical potential $V(r)$ is known from the phenomenological
analyses of elastic scattering data and solve the radial 
Schr\"odinger equation
\begin{equation}
 \frac{d^{2}y_{\ell}(r)}{dr^{2}} + [ k^{2}- U(r) -\frac{\ell(\ell+1)}{r^{2}}]
\:y_{\ell}(r) = 0,
\end{equation}
where $k$ is the wavenumber and $U(r)$ is related to the optical 
potential $V(r)$ through $U(r)=(2m/\hbar^{2})V(r)$.
The solutions for the radial wavefunctions, $y_{\ell}(r)$, are
 normalised according to
\begin{equation}
y_{\ell}(r) = e^{i\delta_{\ell}}[ \cos\delta_{\ell} F_{\ell}(kr) +
\sin\delta_{\ell} G_{\ell}(kr)],
\end{equation}
where $F_{\ell}$ and $G_{\ell}$ are the regular and irregular Coulomb 
functions.
With this normalisation the partial wave amplitudes, $T_{\ell}$, are written 
in terms of the phase-shifts $\delta_{\ell}$ as 
\begin{equation}
T_{\ell}= e^{i\delta_{\ell}} \sin\delta_{\ell}
\end{equation}
The partial wave amplitudes may also be calculated from the relation
\begin{equation}
T_{\ell}=-k^{-1}\int_{0}^{\infty}F_{\ell}(kr)U(r) y_{\ell}(r) dr
\end{equation}

Now we define the optical potential due to breakup as $U_{bu}(r)$ and write the 
full potential $U(r)$ as $(U(r) - U_{bu}(r)) + U_{bu}(r)$. With this
definition, Eq. 1 can be recast as
\begin{equation}
 \frac{d^{2}y_{\ell}(r)}{dr^{2}} + [ k^{2}- (U(r)-U_{bu}(r)) - U_{bu}(r)
 -\frac{\ell(\ell+1)}{r^{2}}]\: y_{\ell}(r) = 0
\end{equation}
A discussion of this  problem may be found in 
Ref.~\cite{RodTha,Wat69}.

By solving the Schr\"odinger equation for the "bare" potential $U(r)-U_{bu}(r)$, 
\begin{equation}
 \frac{d^{2}v_{\ell}(r)}{dr^{2}} + [ k^{2}- (U(r)-U_{bu})
 -\frac{\ell(\ell+1)}{r^{2}}]\: v_{\ell}(r) = 0,
\end{equation}
we obtain the radial wavefunctions $v_{\ell}(r)$, the phase shifts
$\delta'_{\ell}$, and the partial wave amplitudes, $T'_{\ell}$.

The relation between the partial wave amplitudes are 
\begin{equation}
 T_{\ell} = T'_{\ell} +T^{bu}_{\ell}
\end{equation}

The partial wave amplitudes for breakup are thus given by the difference
between the partial wave amplitudes obtained for the potentials 
$U(r)$ and $U(r)-U_{bu}(r)$,
respectively. These can also be calculated from 
the formula (see e.g.~\cite{RodTha,Wat69})

\begin{equation}
T^{bu}_{\ell}=-k^{-1}\int_{0}^{\infty}v_{\ell}(kr)U_{bu}(r) y_{\ell}(r) dr
\end{equation}

It should be noted that the integral in Eq. 8 contains the radial wavefunctions 
$v_{\ell}$, obtained with the bare potential $U(r)-U_{bu}(r)$, as well as 
the radial wavefunctions
$y_{\ell}$, obtained with the full potential $U(r)$. The assumption implicite
therein is that the bare potential cannot lead to the breakup reaction.

The breakup probability for a certain $\ell$-value 
is determined from the partial wave amplitudes 
$T^{bu}_{\ell}$ according to
\begin{equation}
\mathrm{Breakup \ probability} = |T^{bu}_{\ell}|^{2}
\end{equation}

These breakup probabilities are compared with those calculated within the
PFDWBA theory of breakup reactions
(discussed in the next section) to determine the breakup potential,
$U_{bu}$. In our procedure, this potential (with certain ${\it a priori}$
assumed form) is varied to reproduce the breakup probabilities calculated
within PFDWBA. It may be remarked here that when the potential $U_{bu}(r)$ is 
varied, new values of $T'_{\ell}$ have to be calculated from the solution 
of eq. 6, whereas the values of $T_{\ell}$ remain unchanged. 

It is interesting to note that the reaction cross section in the elastic
scattering is known from the solutions of eq. (1). Furthermore the total 
breakup cross section can be calculated from the breakup probabilities.
Even then, the reaction cross section for the bare potential
$U(r)-U_{bu}(r)$, will depend on the shape of the potential,  
$U_{bu}(r)$. 

In Ref.~\cite{Ing96}, the peripheral dominance of the breakup process
made us parametrize the breakup potential as a Gaussian with the addition
of a Woods-Saxon form factor. In this work, however, this preconceived
view is avoided and a Fourier-Bessel (FB) expansion 
\begin{equation}
U_{bu}(r)= \sum_{n=1}^{N} a_{n} \frac{sin(n \pi r)}{R},
\end{equation}
is used for both the real as well as imaginary parts of the breakup potential.
The radius R was kept fixed at 18 fm in all calculations.

\section{Microscopic calculations of breakup probabilities.}

In the theory of the breakup reaction ($d \rightarrow p + n$ ) formulated
within the framework of the post form distorted-wave Born-approximation,
the probability of breakup  $P_{\ell_a}^{b-up(d,p)}$ is defined
by ~\cite{shyam80}
\begin{eqnarray}
\sigma_{total}^{b-up}(d,p) & = & \int d\Omega_p dE_p \frac{d^2 \sigma(d,p)}
                             {d\Omega_p dE_p} \nonumber \\\
                         & = & \frac{\pi}{k_d^2} \sum_{\ell_d} (2\ell_d + 1)
		               P_{\ell_d}^{b-up(d,p)},
\end{eqnarray}
where $\frac{d^2\sigma(d,p)}{d\Omega_p dE_p}$ is the double differential 
cross section for the inclusive breakup reaction (d,p), which is the sum
of the elastic and inelastic breakup modes. These are given by
\begin{eqnarray}
\frac{d^2\sigma(elastic)}{d\Omega_p dE_p} & = & \frac{\mu_d \mu_p \mu_n}
                              {(2\pi)^5 \hbar^6}
                              \frac{k_p k_n}{k_d}
                              \sum_{\ell_n m_n}\mid T_{\ell_n m_n}
                              \mid^2,
\end{eqnarray}
where the T-matrix $T_{\ell_n m_n}$ is  
\begin{eqnarray}
T_{\ell_n m_n} & = &D_{0} \int d^3 r \chi_{p}^{(-)*}({\bf{k_p}}, 
                     \frac{A}{A+1}{\bf{r}})
                     \frac{\chi_{\ell_n}}{k_nr} Y_{\ell_n m_n}(\hat r) 
                     \chi_{d}^{(+)}( {\bf{k_d,r}}) \Lambda(r) P(r)
\end{eqnarray}
$D_{0}$ is the well known zero range constant for the
$d \rightarrow p + n $ vertex.
$\Lambda(r)$ and $P(r)$ are the finite range and nonlocality correction
factors respectively. $\chi^{\pm}$ are the optical model wave functions
in the respective channels with k's being the corresponding wave vectors.
                
The inelastic breakup cross section is given by              
\begin{eqnarray}
\frac{d^2\sigma(inelastic)}{d\Omega_p dE_p} & = & \frac{\mu_d \mu_p \mu_n}
                              {(2\pi)^5 \hbar^6}
                              \frac{k_p k_n}{k_d}
       \sum_{\ell_n m_n}(\sigma_{\ell_n}^{reaction}/\sigma_{\ell_n}^{elastic})
                              \mid T_{\ell_n m_n}-T_{\ell_n m_n}^0 \mid^2
\end{eqnarray}
In this equation $\sigma_{\ell_n}^{reaction}$ and $\sigma_{\ell_n}^{elastic}$
are the reaction and elastic scattering cross sections for the neutron - target    
system corresponding to the partial wave $\ell_n$ respectively. The $T$ matrix
$T_{\ell_n m_n}^{0}$ is defined in the same way as Eq. (13) with the
elastics scattering wave function
$\chi_{\ell_n}$ being replaced by the spherical Bessel function. 
It may be noted
that Eq. (14) includes contributions from all the inelastic
channels of the neutron + target system ( see e.g ~\cite{Shyam84} for complete detail ). 

The angular integration in Eq. (13) is performed by introducing the partial 
wave expansion of the distorted waves and using the orthogonality of the 
spherical harmonics. The resulting slowly converging radial integrals are
evaluated very effectively by following the contour integration technique
of Vincent and Fortune ~\cite{vin70}.

We require the optical potentials in the incident and outgoing channels
as input in our calculations. In the results presented in this paper, the
deuteron optical potentials were taken from the global sets given by
Daehnick, Childs and Vrcelj ~\cite{dae80} whereas the potentials of 
Becchetti and Greenlees ~\cite{becc69} were used in the neutron and
proton channels. 

The total breakup probability $P_{\ell_d}$ is defined as following:
\begin{eqnarray}
P_{\ell_d}^{b-up,d} & = & P_{\ell_d}^{b-up(d,pn)}(elastic) + 
                          P_{\ell_d}^{b-up(d,p)}(inelastic) + \nonumber \\
                    &   &   P_{\ell_d}^{b-up(d,n)}(inelastic)  
\end{eqnarray}                    
In Fig. ~\ref{fig:figa} we show the results for the breakup 
probability for the deuteron incident on a $^{51}$V target at the beam energy
of 56 MeV. We can see that the $(d,p)$ and $(d,n)$ breakup probabilities
are similar in shape and absolute magnitude. The elastic breakup
probability is much smaller and shows a different behaviour as a function
of $\ell_d$.

As discussed in Ref. ~\cite{Shyam84} the total cross section can also 
be expressed as  
\begin{equation}
\sigma_d^{breakup} = 2 \pi \int db\; b\; P_{\ell_{d}}^{b-up,d}
\end{equation}
where the impact parameter, $b$, is related to the angular momentum, $\ell_d$,
and the wavenumber $k_d$, through $b = (\ell_d+ 1/2)/k_d$.

The cross sections for the inelastic (d,p), inelastic (d,n) and  
elastic (d,pn) breakup processes were found to be 290 mb, 214 mb, and 122 mb
respectively. This leads to a total breakup cross section of 627 mb.
It may be noted that our total (d,p) breakup cross section 
(which is the sum of inelastic (d,p) and elastic (d,pn) cross sections)
is 412 mb which is in  
in reasonable agreement with the measured value of 481 mb reported by Matsuoka
et al. ~\cite{Mat80}.

\section{Results and discussions}

The fitting procedure was the same way as that described in Ref. 
~\cite{Ing96}. However, the imaginary part of the breakup potential,
$U_{bu}$, was assumed to be less than that of the full potential $U(r)$, so
that ($U(r)-U_{bu}(r)$) was absorptive for all the radii. The 
errors associated with the calculated breakup
probabilities in the optical model (Eq. (9) ) were about
10$\%$. In the fitting procedure, 14 coefficients ( $a^{n}$ as defined in 
Eq. 10) were varied simultaneously for each potential to get 
a minimum in the $\chi^{2}$. 

In the calculations for ($\alpha-^{3}$He) breakup ~\cite{Ing96} it was found
that the gross properties of the breakup probabilities could be reproduced
with a purely real as well as a purely imaginary breakup potential. An
imaginary part of the 
potential was needed for fitting them at only small $\ell$-values.
However, in case of the deuteron we found it impossible to reproduce 
the breakup probabilities ($P_{\ell_d}$) without a complex breakup potential.
One of the reasons for this could be the fact that the breakup probabilities in the 
this case are considerably larger than those for the
($\alpha-^{3}$He) reaction.  Fig. ~\ref{fig:figb} shows the best fit
obtained with a purely real breakup potential whereas Fig.~\ref{fig:figc} 
with a purely imaginary one. As can be seen from these figures, the imaginary
potential is very important for reproducing the breakup probabilities at
small $\ell$-values, while the real part is required to reproduce these
for the grazing partial waves and beyond. This suppports the observation made 
in ~\cite{Ing96b} that the real potential, due to non eikonal effects,
increases the absorption at large radii and enhances strongly the peripheral
collisions. 

The best fit result obtained with a complex potential is shown in
Fig. ~\ref{fig:figd}.  Both potentials include 14 terms in the FB expansion.
It can be seen that the breakup probabilities calculated within
PFDWBA (shown in  Fig. ~\ref{fig:figa}), are reproduced very well.
The imaginary part of the breakup potential is found to be 
very strong. This is not surprising since the inelastic interactions
(which could cause the breakup of the projectile) are 
taken into account by the imaginary part of $U(r)$. These will automatically 
be included in the imaginary part of $U_{bu}$. In the folding model
calculations of the elastic scattering where the imaginary part of the
potential is phenomenologically determined, it is mainly the real potential  
which is of interest. Therefore, our result suggest that the inclusion 
of the potential, $U_{bu}$, in these calculations will lead to a strong
modification to the real part of the folding model potentials.

Yabana et al ~\cite{Yab92} have investigated the effects of breakup for
the elastic scattering of 80 MeV deuterons from $^{58}$Ni. 
Both real and imaginary parts of the DPP obtained by these authors
are weaker than those obtained in our work. It
may be remarked here that, these authors have neglected the Coulomb 
breakup process, which can be quite large for the deuteron target
interaction~\cite{Baur76, Shyam84}. In our work, on the other hand, this is 
included on the same footing as the nuclear breakup. Moreover, our breakup
probabilities lead to the cross sections which agree well with the
experimental data on the deuteron breakup.  

We should, however, stress that, there are (as in optical model 
calculations) ambiguities in the breakup potentials obtained by us. For         
instance, in Fig. ~\ref{fig:fige} we show the results obtained 
when the number of terms in the real potential is decreased from 14 to 4. 
The fit to the PFDWBA breakup probabilities is
still reasonable. This therefore, makes it difficult to arrive at a 
definite conclusion about the shapes of the DPP. Nevertheless, it is 
interesting to note that the real part of the DPP obtained in this way
is still larger than those of Yabana et al.~\cite{Yab92}

Next we discuss the breakup probability for the elastic breakup ( a 
process in which the target nucleus remains in the ground state).
In theories like the one suggested by Bertsch, Brown and Sagawa
~\cite{Ber89} the reaction cross section in nucleus-nucleus collisions
are calculated from collisions between nucleon-nucleon pairs in 
the projectile and target nuclei. In such approaches it is necessary to
include a correction due to the elastic breakup of the projectile.
We term the corresponding potential as the "dissociation" potential 
($U_{dis}$) to separate it from the potential $U_{bu}$ defined earlier.
It would also be worthwhile to study how the dissociation affects
the optical potential and the reaction cross section. $U_{dis}$ is obtained
by fitting (in the same way as described above) the elastic breakup
probabilites as calculated by PFDWBA theory 
(see Fig. ~\ref{fig:figa}). We stress that these include both 
Coulomb and nuclear breakup as well as their interference terms.

The results obtained with a purely real $U_{dis}$ are shown in
Fig. ~\ref{fig:figf}. The fits to the elastic breakup probabilities are 
satifactory for the grazing partial waves. However, those for the 
lower partial waves are poorer. On the other hand attempts to fit
them with a purely imaginary $U_{dis}$ resulted
in a very bad agreement.  Therefore we reduced the real potential shown 
in Fig. ~\ref{fig:figf} and tried to reproduce the 
data by a variation of the imaginary potential. The results from this
search is shown in Fig. ~\ref{fig:figg} and it is 
evident that the maximum is very badly reproduced.
However, when both the real and imaginary potentials are varied
a good fit to the elastic breakup probabilities are obtained, even if
the shape of the imaginary part of $U_{br}$ so 
obtained looks somewhat unusual. Figure ~\ref{fig:figh}  shows
one of the best fit potentials obtained in this way.

As discussed in section 2, the partial wave
amplitude for the potential $U(r)$ is given by the sum of the partial 
wave amplitudes the potentials,$U(r)-U_{bu}(r)$ and $U_{bu}(r)$.
This is not true for the reaction cross sections calculated from each 
potential separately, since these are sensitive to interference effects.
The reaction cross section for the full potential is 1571 mb.
With the real potential shown in Fig. ~\ref{fig:figf} the reaction
cross section for the potential, $U(r)-U_{bu}(r)$, is found to be 1405 mb.
The difference (166 mb) is somewhat larger than the value (122 mb) obtained
for the total elastic breakup cross section using Eq. (11).
With the complex potential shown in Fig. ~\ref{fig:figh}, the
reaction cross section without breakup potential
does not decrease, instead it goes to a value of 2173 mb.
These examples show the importance of including the effects of the
dissassociation in the optical model. The inteference effects are
important and have to be treated correctly.

Our calculations, therefore, indicate that the dissassociation of the deuteron
should give a substantial contribution to the dynamical polarisation
potential. The assumption that the optical potential, obtained from a fit
to elastic scattering data, include effects of dissassociation and 
inelastic breakup in a correct way must be questioned. 
There are examples, as in Refs. ~\cite{shyam80}, where the absorption due
to breakup is even larger than the total absorption predicted by the optical
model for large $\ell$-values.
Since the elastic breakup is considerably larger than the inelastic breakup
for large $\ell$-values, this indicates that the optical 
models should be modified with a contribution of the dissassociation 
process, which may not necessarily have an effect on the angular
distributions for elastic scattering.  

\section{Conclusion}

In conclusion, our study of the effects of the breakup on the optical 
potential for the elastic scattering of 56 MeV deuterons from $^{51}V$ 
shows that breakup gives a substantial contribution to this potential. 
We found that the strong enhancement of the peripheral collisions requires that
this contribution has a real part. Thus the folding model calculations should
include a dynamical polarisation potential.

The investigations of the effects of the dissassociation also indicate
a contribution to the dynamical polarisation potential. However,
the conventional shapes used for the optical potential do
not account for the dissassociation process in a correct way. Therefore the 
investigation should be repeated with more sophisticated optical model 
potentials with different shapes.   

We plan to calculate breakup probabilities
for all breakup channels for $\alpha$-particles. We believe that these 
together will give total breakup probalities of the same order as those
of the deuterons studied here, at least for energies above 40 MeV/A. 
We see no reason why the breakup potential for $\alpha$-particles
should be purely imaginary when it is complex for deuterons.
Therefore we believe that all light particles require dynamical
polarisation potentials in folding model calculations at higher
energies and that the energy dependence of effective interactions,
which presently reproduce $\alpha$-particle scattering, 
should to be modified.

\begin{figure}
\begin{center}
\end{center}
\caption{Calculated breakup probabilities in the scattering of 56 MeV deuterons
from $^{51}$V.}
\label{fig:figa}
\end{figure}

\begin{figure}
\begin{center}
\caption{Results obtained with a real breakup potential. In the upper part,(a),
 the dashed curve shows the nominal potential, the solid curve the
breakup potential and the dotted curve the difference. In the lower part, (b),
the open circles show the total breakup probabilities from Fig.1 and the solid
circles the fitted values }
\end{center}
\label{fig:figb}
\end{figure}

\begin{figure}
\begin{center}
\end{center}
\caption{Results obtained with an imaginary breakup potential.
In the upper part,(a),
 the dashed curve shows the nominal potential, the solid curve the
breakup potential and the dotted curve the difference. In the lower part, (b),
the open circles show the total breakup probabilities from Fig.1 and the solid
circles the fitted values }
\label{fig:figc}
\end{figure}

\begin{figure}
\begin{center}
\end{center}
\caption{Results obtained with a complex breakup potential.
In the upper parts ,(a) and (b),
 the dashed curve shows the nominal potential, the solid curve the
breakup potential and the dotted curve the difference. In the lower part, (c),
the open circles show the total breakup probabilities from Fig.1 and the solid
circles the fitted values }
\label{fig:figd}
\end{figure}

\begin{figure}
\begin{center}
\end{center}
\caption{Results obtained when the number of terms in the FB-expansion of the
real potential has been decreased from 14 to 4. The notations are the
 same as in Figure 4. }
\label{fig:fige}
\end{figure}

\begin{figure}
\begin{center}
\end{center}
\caption{Results obtained with a real dissociation potential. }
\label{fig:figf}
\end{figure}

\begin{figure}
\begin{center}
\end{center}
\caption{Results obtained with an imaginary dissociation potential.
The real potential is in this case 50 \% of the real potentiatial shown in
Figure 6 }
\label{fig:figg}
\end{figure}

\begin{figure}
\begin{center}
\end{center}
\caption{Results obtained with a complex dissociation potential. }
\label{fig:figh}
\end{figure}

\end{document}